\documentclass[prl,twocolumn,showpacs,superscriptaddress]{revtex4}
\usepackage{graphicx,bm,amssymb,amsmath,slashed}

\newcommand{\Eq}[1]{Eq.~\eqref{#1}}
\newcommand{\eq}[1]{\eqref{#1}}

\newcommand{\beq}{\begin{equation}}
\newcommand{\eeq}{\end{equation}}

\newcommand{\beqa}{\begin{eqnarray}}
\newcommand{\eeqa}{\end{eqnarray}}

\newcommand{\Beqa}{\begin{eqnarray*}}
\newcommand{\Eeqa}{\end{eqnarray*}}

\newcommand{\nn}{\nonumber}

\newcommand{\bra}[1]{\left\langle{#1}\right\rvert}
\newcommand{\ket}[1]{\left\lvert{#1}\right\rangle}

\newcommand{\vect}[1]{\mathbf{#1}}

\DeclareMathOperator{\im}{Im}
\DeclareMathOperator{\re}{Re}

\newcommand{\up}{\uparrow}
\newcommand{\down}{\downarrow}

\begin{document}
\newcommand{\cI}{{\mathcal I}}
\title{A Universal Energy Functional for Trapped Fermi Gases with Large Scattering Length}

\author{Shina Tan}
\affiliation{School of Physics, Georgia Institute of Technology, Atlanta, Georgia 30332, USA}

\begin{abstract}
Yoram Alhassid conjectured that the total energy of a harmonically trapped two-component Fermi gas with large scattering length
is a linear functional of the occupation probabilities of single-particle energy eigenstates.
We confirm his conjecture and derive the functional explicitly.
We show that the functional applies to \emph{all smooth potentials} having a minimum, not just harmonic traps.
We also calculate the occupation probabilities of high energy states.
\end{abstract}
\maketitle

\textbf{Introduction--}
It is well known that the energy of a noninteracting
Fermi gas in a trap can be expressed as a summation over single-particle energy eigenstates
\beq\label{E noninteracting}
E=\sum_{\nu\sigma}\epsilon_\nu n_{\nu\sigma},
\eeq
where $\nu$ and $\sigma$ label the orbital and spin states respectively,
$\epsilon_\nu$ is the single-particle energy level, and
$n_{\nu\sigma}$ is the occupation probability of the state $(\nu,\sigma)$.

For interacting systems, however,
there is no general relation between the total energy, which includes interaction energy, and $n_{\nu\sigma}$ alone.

In this Letter we study a remarkable exception to this general rule,
in a Fermi gas with \emph{strong} interactions.

We consider a two-component ($\sigma=\up, \down$)
Fermi gas in which the $s$-wave scattering length $a$ and other relevant length scales,
such as the mean inter-particle spacing $d$ and the thermal de Broglie wave length,
are all much larger than the range of the interaction $r_e\to0$.
Applications of this $s$-wave contact interaction model range from
ultracold thin vapors of neutral atoms near broad Feshbach
resonances to the neutron gas with density and temperature below
nuclear scales, or from extremely cold ($10^{-9}$-$10^{-6}$K) to extremely hot ($10^9$K) matter.

The energy of such a Fermi gas was found to be \cite{TanEnergetics, Braaten2008Mar, Combescot2009Jan}
\begin{align}\label{E old}
E&=\frac{\hbar^2\cI}{4\pi ma}+\int\frac{d^3k}{(2\pi)^3}\sum_\sigma\frac{\hbar^2k^2}{2m}\Big(\rho_{\vect k\sigma}-\frac{\cI}{k^4}\Big)\nn\\
&\quad+\int d^3r\sum_\sigma n_\sigma(\vect r)V(\vect r),
\end{align}
where $\rho_{\vect k\sigma}$ is the momentum distribution,
$n_\sigma(\vect r)$ is the spatial density distribution,  $m$ is  each fermion's mass, $V(\vect r)$ is the external potential, and
$\cI=\lim_{k\to\infty}k^4\rho_{\vect k\sigma}$
is the \emph{contact}, a parameter characterizing the number of small pairs of fermions \cite{TanEnergetics, Braaten2008Mar}.
The contact is at the center of many universal relations for fermions with $s$-wave contact interaction
\cite{TanEnergetics, Tan2D, Zwerger2007Jul, Baym2007Jul,
Braaten2008Mar, Werner2008Mar, Braaten2008Jun, Werner2008Jul, ZhangLeggett2008Sep, Strinati2008Nov,
Combescot2009Jan, Schneider2009Mar, Zwerger2009Apr, Baym2009May,
Blume2009Sep, Randeria2009Oct, Su2009Dec, Werner2010Jan, Braaten2010Jan, Randeria2010Feb, Son2010Feb, Hu2010Mar, Strinati2010May,
Zwerger2010Jul, Braaten2010Aug, DeSilva2010Dec, Carlson2010Dec, Drut2010Dec, Goldberger2010Dec, Zwerger2011Jan,
Hulet2005May, Vale2010Jan, Jin2010Feb, Salomon2010Apr}.

Equation~\eq{E old}, containing \emph{two continuous distributions} [$\rho_{\vect k\sigma}$ and $n_{\sigma}(\vect r)$],
is considerably more complicated than \Eq{E noninteracting} which involves merely the occupation of single-particle levels.
In the case of a harmonic trap, $n_{\nu\sigma}$ are just a \emph{discrete} set of numbers.

Yoram Alhassid conjectured that for harmonically trapped fermions with large scattering length,
the total energy might still be a linear functional of $n_{\nu\sigma}$ \cite{Alhassid}.

We will show that his conjecture is true by deriving the functional explicitly.
We also show that it is valid for \emph{all smooth potentials} having a minimum (which we set to
zero without loss of generality).
One can equally well apply this functional to anharmonic traps, periodic potentials 
(eg, an optical lattice for cold atoms), or disordered potentials, etc.

This universal energy functional is
\beq\label{generalized}
E=\frac{\hbar^2\mathcal I}{4\pi ma}+\lim_{\epsilon_\text{max}\to\infty}\Big(
\sum_{\epsilon_\nu<\epsilon_\text{max}}\epsilon_\nu n_\nu-\frac{\hbar\,\mathcal I}{\pi^2}\sqrt\frac{\epsilon_\text{max}}{2m}\Big),
\eeq
where $n_\nu\equiv\sum_\sigma n_{\nu\sigma}$.
The contact $\cI$ is contained in the asymptotic behavior of
\beq\label{rho sigma def}
\rho_\sigma(\epsilon)\equiv\sum_{\nu}n_{\nu\sigma}\delta(\epsilon-\epsilon_\nu)
\eeq
at high energy,
\beq\label{asymp rho}
\rho_\sigma(\epsilon)\big|_\text{coarse grained}=\frac{\hbar\mathcal I}{4\pi^2\sqrt{2m}}\epsilon^{-3/2}+O(\epsilon^{-5/2}).
\eeq

We also calculate the occupation probabilities of individual single-particle energy eigenstates at
$
\epsilon_\nu\gg\max\{|E|/N,\,\hbar^2/ma^2,\,\hbar^2/md^2,\,\Delta V\},
$
where $N$ is the number of fermions, and $\Delta V$ is the characteristic range of potentials involved in the $N$-body state.
(For example, if many fermions form a cloud in a trap, $\Delta V$ is the change of $V(\vect r)$ from the trap minimum
to the edge of the cloud.)
The result is
\begin{align}\label{asymp n}
n_{\nu\sigma}&=\frac{1}{k_\nu^4}\int C(\vect r)|\phi_\nu(\vect r)|^2d^3r+\frac{4m^2}{\hbar^2k_\nu^6}\int\vect D(\vect r)\cdot\vect j_\nu(\vect r)d^3r\nn\\
&\quad+O(\epsilon_\nu^{-3}),
\end{align}
where $k_\nu\equiv{\sqrt{2m\epsilon_\nu}}/\hbar$, $C(\vect r)$ is the contact density \cite{TanEnergetics, Braaten2008Mar}
[related to the total contact: $\int C(\vect r)d^3r=\cI\,$],
$\phi_\nu(\vect r)$ is the normalized wave function of the $\nu$-th single particle orbital, satisfying the Schr\"{o}dinger equation
\beq\label{1body Schrodinger}
\big[-(\hbar^2/2m)\nabla^2+V(\vect r)\big]\phi_\nu(\vect r)=\epsilon_\nu\phi_\nu(\vect r),
\eeq
$\vect D(\vect r)$ is the ``contact current" [see \Eq{D def} below],
and $\vect j_\nu(\vect r)\equiv(\hbar/m)\im\phi_\nu^*\nabla\phi_\nu\sim O[(\epsilon_\nu/m)^{1/2}|\phi_\nu|^2]$ is the probability current
of the $\nu$-th single-particle orbital state.

Equations~\eqref{generalized}, \eqref{asymp rho} and \eqref{asymp n} apply to
both energy eigenstates and thermal ensembles, 
both equilibrium and non-equilibrium states, both few-body and many-body systems, both strong ($|a|\gtrsim d$) and weak ($|a|\ll d$) interactions,
both symmetric ($N_\up=N_\down$) and polarized ($N_\up\ne N_\down$) states.
In the many-body regime, they are valid for \emph{all phases}, including normal and superfluid phases.

%
%

In the following we first derive an expansion for the one-particle density matrix $p_\sigma(\vect r,\vect r+\vect b)\equiv\langle
\psi_\sigma^\dagger(\vect r)\psi_\sigma(\vect r+\vect b)\rangle$
at a small separation $\textbf b$. From this we derive \Eq{generalized}
(exploiting the propagator of a single particle in a short imaginary time),
\Eq{asymp rho}, and \Eq{asymp n}. The derivations are for energy eigenstates but can be easily
extended to thermal ensembles and non-equilibrium states.


\textbf{One-Particle Density Matrix--}
Consider a normalized $N$-body energy eigenstate ($N=N_\up+N_\down$)
\begin{align}
\ket{\phi}=&(N_\up!\,N_\down!)^{-1/2}\int D^\up_1D^\down_1
\phi(\vect r_1,\cdots,\vect r_{N_\up},\vect s_1,\cdots,\vect s_{N_\down})\nn\\
&\times\psi_\up^\dagger(\vect r_1)\cdots\psi_\up^\dagger(\vect r_{N_\up})\psi_\down^\dagger(\vect s_1)\cdots\psi_\down^\dagger(\vect s_{N_\down})\ket{0},
\end{align}
where $\ket{0}$ is the particle vacuum, $\psi_\sigma^\dagger(\vect r)$ is the standard fermion creation operator,
and we have introduced short-hand notations
$D^\up_i\equiv\prod_{\mu=i}^{N_\up}d^3r_\mu$ and
$D^\down_i\equiv\prod_{\mu=i}^{N_\down}d^3s_\mu.$
When $\vect r_1$ and $\vect s_1$ are close,
$\phi(\vect r_1,\cdots,\vect r_{N_\up},\vect s_1,\cdots,\vect s_{N_\down})$ satisfies the Bethe-Peierls boundary condition
\begin{align}\label{boundary}
\phi=&A\big(\tfrac12(\vect r_1+\vect s_1);\vect r_2\cdots\vect r_{N_\up}\vect s_2\cdots\vect s_{N_\down}\big)\big(|\vect r_1-\vect s_1|^{-1}-a^{-1}\big)\nn\\
&+O(|\vect r_1-\vect s_1|).
\end{align}

We now expand the 1-particle density matrix
\begin{align}
&p_\up(\vect r,\vect r+\vect b)=\bra{\phi}\psi_\up^\dagger(\vect r)\psi_\up(\vect r+\vect b)\ket{\phi}\nn\\
&=N_\up\int D^\up_2D^\down_1\,
\phi^*(\vect r,\vect r_2\cdots\vect r_{N_\up}\vect s_1\cdots\vect s_{N_\down})
\phi(\vect r+\vect b,\vect r_2\cdots)
\end{align}
through order $O(b^2)$ at a small distance $b$.
Because of the singularity of $\phi$
when two fermions in different spin states are close (see above),
we divide the $3(N_\up+N_\down-1)$ dimensional integration domain into region $\mathcal{R}_\epsilon$ (in which \emph{every} spin-down fermion
lies outside of the sphere of radius $\epsilon$ centered at $\vect r$, namely $|\vect s_\mu-\vect r|>\epsilon$ for $\mu=1,\cdots,N_\down$)
and its complement, $\overline{\mathcal{R}_\epsilon}$. Here $\epsilon$ is small but $\epsilon>b$.
In $\mathcal{R}_\epsilon$ we expand $\phi(\vect r+\vect b,\vect r_2\cdots\vect r_{N_\up}\vect s_1\cdots\vect s_{N_\down})$ in powers of $\vect b$,
while in $\overline{\mathcal{R}_\epsilon}$ we use \Eq{boundary} which is sufficient for evaluating the integral in $\overline{\mathcal{R}_\epsilon}$ through order $b^2$.
In $\overline{\mathcal{R}_\epsilon}$ it is possible for two or more spin-down fermions to come inside the small sphere of radius $\epsilon$ centered at $\vect r$,
but the contributions from such cases are suppressed by Fermi statistics and are of higher order than $O(b^3)$.
When the integrals in the two regions $\mathcal{R}_\epsilon$ and $\overline{\mathcal{R}_\epsilon}$ are added,
all dependencies on $\epsilon$ are canceled, yielding the following clean expansion:
\begin{align}\label{rho up expand}
p_\up(\vect r,\vect r+\vect b)&=n_\up(\vect r)+C(\vect r)(-b/8\pi+b^2/24\pi a)+\vect b\cdot\vect u_\up(\vect r)\nn\\
&\quad-3\pi b\vect b\cdot\vect w(\vect r)/2-\pi b{\vect b}\cdot\vect w^*(\vect r)/2\nn\\
&\quad+\sum_{i,j=1}^3v_{\up ij}(\vect r)b_ib_j/2+O(b^3),
\end{align}
where $b_i$ is the $i$-th Cartesian component of $\vect b$,
\beq
n_\up(\vect r)=N_\up\int D^\up_2D^\down_1|\phi(\vect r\vect r_2\cdots\vect r_{N_\up}\vect s_1\cdots\vect s_{N_\down})|^2
\eeq
is the density of spin-up fermions at position $\vect r$,
\begin{align}
C(\vect r)=16\pi^2N_\up N_\down\int D^{\up}_2D^{\down}_2
|A(\vect r;\vect r_2\cdots\vect r_{N_\up}\vect s_2\cdots\vect s_{N_\down})|^2
\end{align}
is the contact density \cite{TanEnergetics, Braaten2008Mar},
\begin{align}
\vect w(\vect r)&\equiv N_\up N_\down\int D^\up_2D^\down_2A^*(\vect r;\vect r_2\cdots\vect r_{N_\up}\vect s_2\cdots\vect s_{N_\down})\nn\\
&\mspace{80mu}\times\nabla_rA(\vect r;\vect r_2\cdots\vect r_{N_\up}\vect s_2\cdots\vect s_{N_\down})
\end{align}
is related to the center-of-mass motion of small pairs,
and
\begin{align}
\vect u_\up(\vect r)&\equiv N_\up\lim_{\eta\to0}\int_{\mathcal{R}_\eta}
D^\up_2D^\down_1\phi^*(\vect r\vect r_2\cdots\vect r_{N_\up}\vect s_1\cdots\vect s_{N_\down})\nn\\
&\mspace{110mu}\times\nabla_r\phi(\vect r\vect r_2\cdots\vect r_{N_\up}\vect s_1\cdots\vect s_{N_\down}),
\end{align}
\begin{align}
v_{\up ij}(\vect r)&\equiv N_\up\lim_{\eta\to0}\int_{\mathcal{R}_\eta}
D^\up_2D^\down_1\phi^*(\vect r\vect r_2\cdots\vect r_{N_\up}\vect s_1\cdots\vect s_{N_\down})\nn\\
&\mspace{70mu}\times\frac{\partial^2}{\partial r_i\partial r_j}
\phi(\vect r\vect r_2\cdots\vect r_{N_\up}\vect s_1\cdots\vect s_{N_\down}).
\end{align}
There is of course a completely analogous expansion for $p_\down(\vect r,\vect r+\vect b)$
involving the same $C(\vect r)$ and $\vect w(\vect r)$.

In addition to the contact density $C(\vect r)$ \cite{TanEnergetics, Braaten2008Mar}, we introduce a ``\emph{contact current}"
\beq\label{D def}
\vect D(\vect r)\equiv(8\pi^2\hbar/m)\im\vect w(\vect r).
\eeq
There is in general no continuity relation between the contact density and the contact current,
because the small pairs may dissociate or associate.

\textbf{Universal Energy Functional--} For any $(N_\up+N_\down)$-body
energy eigenstate $\ket{\phi}$ and any $\beta$ satisfying $\re\beta\ge0$ we define an absolutely convergent series:
\beq\label{J def}
J_\sigma(\beta)\equiv\sum_\nu n_{\nu\sigma}e^{-\beta\epsilon_\nu}=\sum_\nu \bra{\phi} c_{\nu\sigma}^\dagger c_{\nu\sigma}\ket{\phi}
e^{-\beta\epsilon_\nu}.
\eeq
Since the fermion annihilation operator
\beq\label{c expand}
c_{\nu\sigma}=\int d^3r\,\phi_\nu^*(\vect r)\psi_\sigma(\vect r),
\eeq
we have
\beq\label{J int}
J_\sigma(\beta)=\int d^3rd^3r'U_\beta(\vect r,\vect r')p_\sigma(\vect r,\vect r'),
\eeq
where $U_\beta(\vect r,\vect r')\equiv\sum_\nu e^{-\beta\epsilon_\nu}\phi_\nu(\vect r)\phi_\nu^*(\vect r')$
is the propagator of a single particle moving in the potential $V(\vect r)$
within a time $-i\hbar\beta$. For a small positive $\beta$, at $|\vect r-\vect r'|\gg\hbar\sqrt{\beta/m}$
the propagator is exponentially suppressed, while at $|\vect r-\vect r'|\sim\hbar\sqrt{\beta/m}$ we have
a ``short imaginary-time expansion"
\begin{align}\label{U expand}
U_\beta(\vect r,\vect r')&=(2\pi\hbar^2\beta/m)^{-3/2}\big\{1-[V(\vect r)+V(\vect r')]\beta/2\}\nn\\
&\mspace{10mu}\times\exp\big[-m(\vect r-\vect r')^2/2\hbar^2\beta\big]+O(\beta^{1/2}),
\end{align}
provided that $V(\vect r)$ is smooth.
But when $|\vect r-\vect r'|$ is small we also have a systematic expansion
for $p_\sigma(\vect r,\vect r')$ [see above]. Substituting both expansions into \Eq{J int} we obtain
a systematic expansion for $J_\sigma(\beta)$ at small $\beta$:
\begin{align}\label{J sigma expand}
J_\sigma(\beta)&=N_\sigma-(\hbar\cI/4\pi^2)\sqrt{2\pi\beta/m}+\hbar^2\cI\beta/8\pi ma\nn\\
&\quad-\beta\int d^3r V(\vect r)n_\sigma(\vect r)\nn\\
&\quad+(\hbar^2\beta/2m)\int d^3r\sum_{i=1}^3v_{\sigma ii}(\vect r)+O(\beta^{3/2}).
\end{align}
From the $N$-body Schr\"{o}dinger equation
\begin{align}
&\Big\{\sum_{\mu=1}^{N_\up}\Big[\!-\!\frac{\hbar^2}{2m}\nabla_{r_\mu}^2+V(\vect r_\mu)\Big]
+\sum_{\mu'=1}^{N_\down}\Big[\!-\!\frac{\hbar^2}{2m}\nabla_{s_{\mu'}}^2+V(\vect s_{\mu'})\Big]\Big\}\phi\nn\\
&=E\phi,~~~~~\text{if $\vect r_\mu\ne\vect s_{\mu'}$ for all $\mu,\mu'$},
\end{align}
one can show that
\beq
\sum_\sigma\int d^3r\Big[V(\vect r)n_\sigma(\vect r)-\frac{\hbar^2}{2m}\sum_{i=1}^3v_{\sigma ii}(\vect r)\Big]=E,
\eeq
so the summation of \Eq{J sigma expand} over $\sigma$ yields
\beq
\sum_{\nu\sigma}n_{\nu\sigma}e^{-\beta\epsilon_\nu}=N-\frac{\hbar\cI}{2\pi^2}\sqrt\frac{2\pi\beta}{m}
+\frac{\hbar^2\cI\beta}{4\pi ma}-\beta E+O(\beta^{3/2}).
\eeq
Applying $\frac{d}{d\beta}$ to the above expansion,
defining $\rho(\epsilon)=\sum_{\nu\sigma}n_{\nu\sigma}\delta(\epsilon-\epsilon_\nu)$, and taking $\beta\to0$, we find
\begin{align}
E&=\frac{\hbar^2\cI}{4\pi ma}+\lim_{\beta\to0}
\int_0^\infty\Big[\rho(\epsilon)-\frac{\hbar\cI}{2\pi^2\sqrt{2m}}\epsilon^{-3/2}\Big]\epsilon e^{-\beta\epsilon} d\epsilon\nn\\
&=\frac{\hbar^2\cI}{4\pi ma}+
\int_0^\infty\Big[\rho(\epsilon)-\frac{\hbar\cI}{2\pi^2\sqrt{2m}}\epsilon^{-3/2}\Big]\epsilon d\epsilon,
\end{align}
and thus \Eq{generalized}.

\textbf{Asymptotics of $\rho_\sigma(\epsilon)$--}
When $\beta=it/\hbar$ is purely imaginary and the real ``time" $t\to\pm0$,
the net contribution to the integral in \Eq{J int} from
$|\vect r-\vect r'|\gg\sqrt{\hbar|t|/m}$ is exponentially small because of the rapid oscillation of
the propagator $U$. When $|\vect r-\vect r'|\sim\sqrt{\hbar|t|/m}$ the expansion in \Eq{U expand}
with $\beta$ replaced by $it/\hbar$ holds \cite{U}.
Therefore, $J_\sigma(it/\hbar)$ has a ``short real-time expansion" by simply
setting $\beta=it/\hbar$ in \Eq{J sigma expand}, and thus has an $O(\sqrt{|t|})$ singularity at $t=0$.
So the function $\rho_\sigma(\epsilon)$ defined in \Eq{rho sigma def}, which is the Fourier transform
of $J_\sigma$:
\beq
\rho_\sigma(\epsilon)=(2\pi\hbar)^{-1}\int_{-\infty}^\infty J_\sigma(it/\hbar)e^{i\epsilon t/\hbar}dt,
\eeq
has a ``coarse-grained" asymptotic formula at high energy shown in \Eq{asymp rho}.
For a deep trap, the fact that $\rho_\sigma(\epsilon)$ remains a discrete sum of delta functions at large $\epsilon$,
rather than turning into a continuous curve, can be traced to the singularities of $J_\sigma(it/\hbar)$ at nonzero
$t$'s. But in a ``coarse grained" distribution,
$
\rho_\sigma(\epsilon)|_\text{coarse grained}=\int_{-\infty}^\infty g(\epsilon')\rho_\sigma(\epsilon-\epsilon')d\epsilon',
$
where the convolution factor may be chosen as $g(\epsilon')=\exp(-\epsilon'^2/\lambda^2)/(\lambda\sqrt\pi)$ with a large width $\lambda$,
these singularities are ``washed out",
because ${\rho}_\sigma(\epsilon)|_\text{coarse grained}$ is the Fourier transform of the product of $\widetilde{g}(t)$ and $J_\sigma(it/\hbar)$,
where $\widetilde{g}(t)$ is the inverse Fourier transform of $g(\epsilon)$ and decays exponentially at $|t|\gg\hbar/\lambda$.
For the validity of \Eq{asymp rho}, the energy resolution $\lambda$ should not grow faster than constant$\times\sqrt{\epsilon}$.

\textbf{Asymptotics of $n_{\nu\sigma}$--} From \Eq{c expand} we find
\beq
n_{\nu\sigma}=\int d^3r \phi_\nu(\vect r)\int d^3b\,\phi_\nu^*(\vect r+\vect b)p_\sigma(\vect r,\vect r+\vect b).
\eeq
At large $\epsilon_\nu$, the integrand as a function of $\vect b$ oscillates rapidly.
The only significant contribution comes from the power-law singularities of $p_\sigma$ at $\vect b\to0$.
According to \Eq{rho up expand}, the leading order singular term is $\propto|\vect b|$, for which we write
$\phi_\nu^*(\vect r+\vect b)\doteq(-\hbar^2/2m\epsilon_\nu)^2\nabla_b^4\phi_\nu^*(\vect r+\vect b)$ with relative error $\sim O(\epsilon_\nu^{-1})$
according to \Eq{1body Schrodinger} \cite{finiteV}. Integration by parts over $\vect b$ yields $\propto\int d^3b\phi_\nu^*(\vect r+\vect b)\delta(\vect b)$,
leading to the first term on the right hand side of  \Eq{asymp n}.

The next order singular term in $p_\sigma$ is $\propto b\vect b$, for which we write
$\phi_\nu^*(\vect r+\vect b)\doteq(-\hbar^2/2m\epsilon_\nu)^3\nabla_b^6\phi_\nu^*(\vect r+\vect b)$ with relative error $\sim O(\epsilon_\nu^{-1})$.
Integration by parts over $\vect b$ yields $\propto\int d^3b\phi_\nu^*(\vect r+\vect b)\nabla_b\delta(\vect b)=-\nabla\phi_\nu^*(\vect r)$.
Further integrating by parts over $\vect r$, omitting contributions $\sim O(\epsilon_\nu^{-3})$,
and using \Eq{D def} and the identities
$\re\vect w(\vect r)=\nabla C(\vect r)/32\pi^2$ and $\nabla\cdot\vect j_\nu(\vect r)=0$, we obtain
the second term on the right hand side of  \Eq{asymp n}.

Because any single-particle orbital state and its time reversal have the same energy but opposite probability currents,
the second term on the right hand side of \Eq{asymp n} has no net contribution to the distribution $\rho_\sigma(\epsilon)$.

We now illustrate \Eq{asymp n} with a symmetric unitary ($|k_Fa|\gg1$) Fermi gas at zero temperature, confined by
a spherical harmonic trap of angular frequency $\omega$. At large $N$ the local density approximation (LDA) for
the contact density is valid: $C(r)=C_1k_{F0}^4(1-r^2/R^2)^2$. Here $k_{F0}$ is the local Fermi wave number at the trap center, $R$ is the LDA cloud radius,
and $C_1k_F^4$ is the contact density of the homogeneous unitary Fermi gas with Fermi wave number $k_F$.
For a high energy orbital $\nu=(j,l,m_z)$ with energy $\epsilon_\nu=(j+3/2)\hbar\omega$, orbital angular momentum quantum number $l$, and
magnetic quantum number $m_z$ we find \cite{averageC}
\beq
n_{\nu\sigma}\doteq\frac{16C_1N^{5/6}}{5\pi\, 3^{1/6}\xi^{3/4}j^{5/2}}\Big[1-\frac{l(l+1)}{4\sqrt\xi(3N)^{1/3}j}\Big]^{5/2}
\eeq
if $l(l+1)<4\sqrt\xi(3N)^{1/3}j$. For higher $l$, the classical orbit is outside of the LDA cloud radius, and $n_{\nu\sigma}$ becomes exponentially
small. Here $\xi$ is the Bertsch parameter \cite{Baker1999May}, ie the ratio between the ground state energy of the unitary Fermi gas
and that of the noninteracting Fermi gas at the same density. 
In latest numerical and experimental studies, $\xi\lesssim0.38$ \cite{Forbes2010Nov} and $C_1\approx0.12$
\cite{Vale2010Jan, Jin2010Feb, Salomon2010Apr, Stringari2006Apr}.

To conclude, we have shown that the total energy of fermions with large scattering length ($|a|\gg r_e$) in any smooth potential
is a simple linear functional of the occupation probabilities of single-particle energy eigenstates. We have also derived asymptotic expressions
for the occupation probabilities of high energy states.
These results can be verified experimentally by measuring the energy and the occupation probabilities independently.
They also provide robust constraints on theories of trapped Fermi gases, including fermionic atoms in optical lattices.
These results can be extended to lower dimensions, to fermions with unequal masses, and to bosons and Bose-Fermi mixtures.

The universal energy functional \Eq{generalized} implies a new approach to the difficult many-body problem at large scattering length:
by identifying nontrivial constraints on the occupation probabilities, one can
minimize the functional to find the many-body ground state energies in external potentials.


The author thanks Yoram Alhassid, Tin-Lun Ho, Yusuke Nishida, Kenneth O'Hara, Mohit Randeria, Junliang Song, Zhenghua Yu, Hui Zhai and
Fei Zhou for valuable discussions, and the Aspen Center for Physics where this work was finalized.




\begin{thebibliography}{41}
\bibitem[{U()}]{U}
\bibinfo{note}{For certain potentials with steep edges, such as $V(\vect
  r)\propto x^{2s}+y^{2s}+z^{2s}$ (where $s=2,3,4,\cdots$), there are
  additional, highly oscillatory, contributions to $U$ at small $t$ and $|\vect
  r-\vect r'|$ from energetic reflections by the edges. Their contributions to
  $J_\sigma(it/\hbar)$ are of higher order than $O(|t|^{3/2})$.}
\bibitem[{fin()}]{finiteV}
\bibinfo{note}{Although in the entire space $V(\vect r)$ may not have an upper
  bound, in the region where $p_\sigma(\vect r,\vect r+\vect b)$ is \emph{not}
  exponentially small, $V(\vect r+\vect b)$ is finite and hence $V(\vect
  r+\vect b)/\epsilon_\nu\to0$ when $\epsilon_\nu\to\infty$.}
\end{thebibliography}
\end{document}